\documentclass{IEEEtran}
\usepackage{amsmath,amssymb,amsfonts}
\usepackage{multirow}
\usepackage{tikz,pgfplots}
\usepgfplotslibrary{statistics}
\usepackage{siunitx}
\DeclareSIUnit[per-mode=symbol]{\rocauc}{ROC\,AUC}
\usepackage{mathrsfs}%
\usepackage{xcolor}%
\usepackage{textcomp}%
\usepackage{manyfoot}%
\usepackage{booktabs}%
\usepackage{multirow}  
\usepackage{algorithm}%
\usepackage{algpseudocode}%
\usepackage{etoolbox}
\AtBeginEnvironment{table}{\footnotesize}

\usepackage{listings}%
\usepackage{tikz}
\usepackage{caption}
\usepackage{subcaption}
\usepackage{graphicx}
\usepackage{cite}
\usepackage{textcomp}

\graphicspath{}
\pgfplotsset{compat=1.18}

\newcommand{\fT}{\mathbf{f}_\mathrm{T}}

\definecolor{Baseline}{HTML}{FF6361}
\definecolor{Tuned}{HTML}{A1D99B}

\begin{document}
\title{Optimizing In Vivo Oral Lesion Classification from Electrical Impedance Spectroscopy Using Data-driven Approaches}
\author{
Sophie A. Lloyd\textsuperscript{*}, \IEEEmembership{Member, IEEE},
Jacob P. Thönes\textsuperscript{*}, \IEEEmembership{Member, IEEE},
Safina S. Suratwala,
Noor Zaghlula,
Liang Lu, 
Joseph Paydarfar,
Ethan K. Murphy, \IEEEmembership{Member, IEEE},
Sascha Spors, \IEEEmembership{Member, IEEE},
Ryan J. Halter, \IEEEmembership{Member, IEEE}
\thanks{
This work was supported by the NSF Research Traineeship, Transformative Research and Graduate Education in Sensor Science, Technology and Innovation (DGE- 2125733) and NIH’s NIDCR under Award Number 1R21DE031095-01 and NCI under 1R01CA237654-01A1.  
Jacob P. Thönes received funding from the Deutsche Forschungsgemeinschaft (DFG, German Research Foundation) – SFB 1270/4 - 299150580.
}
\thanks{
\textsuperscript{*} These corresponding authors contributed equally.
Sophie A. Lloyd (e-mail: sophie.lloyd.th@dartmouth.edu),
Ethan K. Murphy, Safina S. Suratwala, and Ryan J. Halter are with Thayer School of Engineering at Dartmouth College, Hanover, NH, USA. 
Noor Zaghlula, Ryan J Halter, and Joseph Paydarfar are with Geisel School of Medicine, Dartmouth College, Hanover, NH, USA. 
Liang Lu and Joseph A. Paydarfar are with Dartmouth Hitchcock Medical Center, Lebanon, NH, USA. 
Jacob P. Thönes (e-mail: jacob.thoenes@uni-rostock.de) and Sascha Spors are with Institute of Communications Engineering, University of Rostock, Rostock, 18055, Germany.}
\thanks{
This work has been submitted to the IEEE for possible publication. Copyright may be transferred without notice, after which this version may no longer be accessible.}
}

\maketitle
\begin{abstract}
Oral cancer is a significant global health burden, and early detection remains a critical clinical need. 
Electrical impedance spectroscopy (EIS) offers a promising non-invasive approach for real-time tissue characterization, but classification frameworks that jointly leverage multiple impedance features for in vivo oral lesion discrimination remain underdeveloped.
This paper presents a machine-learning (ML) pipeline to optimize classification of in vivo oral pathology from EIS data collected using a handheld, bedside device.
Impedance measurements were acquired from 104 patients undergoing oral cancer resection or biopsy. 
Three classification tasks were evaluated: (1) healthy vs. cancer, (2) multi-class lesion-type discrimination (cancer, high-grade dysplasia, non-malignant), and (3) multi-class discrimination between the three lesion pathologies and healthy tissue.
For each task, signal frequencies were independently ranked and reduced using PCA, and different current injection/voltage measurement (IIVV) pattern geometries were tested. 
Classification performance was assessed through leave-one-patient-group-out cross-validation to ensure robustness on unseen patients.
Input data dimensionality was reduced by up to 99\% across all tasks while improving diagnostic accuracy over baseline models trained on the full dataset.
A logistic regression model achieved the highest binary classification accuracy of 80\% with an AUC of 0.90, while multi-class scenarios maintained AUCs above 0.82.
All top-performing models utilized the significantly reduced IIVV set as input. 
The proposed pipeline advances EIS-based cancer detection by providing a robust, computationally efficient, and clinically practical framework for early diagnosis of oral cancer lesions, with a methodology readily generalizable to other EIS devices and applications.
\end{abstract}

\begin{IEEEkeywords}
Electrical impedance spectroscopy, oral cancer,
machine learning (ML) classification, dimension reduction
\end{IEEEkeywords}

\section{Introduction}
\label{Introduction}


Oral squamous cell carcinoma (OSCC) is a major global health problem: in \num{2025}, there were an estimated \num{60000} new cases and \num{13000} deaths in the United States \cite{siegel_cancer_2025}. 
Late-stage diagnosis is common for oral cancer and requires more aggressive treatment, leading to higher mortality, morbidity, and health-care costs. 
Early detection substantially improves outcomes: reported five‑year survival rates increase from about \SI{58}{\percent} for late-stage disease to roughly \SI{89}{\percent} when diagnosed early \cite{NCI_Cancer_Oral_Cavity}. 
Despite this, frontline screening and diagnosis of early-stage OSCC and premalignant lesions remain inadequate.

The clinical standard for diagnosis is an invasive tissue biopsy with histopathological analysis.
While histopathology is highly accurate for distinguishing malignant from non-malignant tissue \cite{speight_pathology_2018}, grading of dysplasia shows substantial inter-observer variability \cite{kujan_why_2007,speight_interobserver_2015}. 
Biopsies are also invasive and resource intensive, requiring specialized equipment and trained personnel. 
Many dental clinics do not have ready access to histopathology services, and only about \SI{25}{\percent} of dentists perform oral biopsies, commonly citing costs, distance to laboratories, and lack of training \cite{cassol_spanemberg_experiences_2023}. 
These limitations reduce opportunities for timely diagnosis and early intervention.

Non-invasive diagnostic technologies are therefore needed to enable earlier detection of OSCC and improve risk stratification of pre-malignant lesions across diverse clinical settings.
Optical modalities, such as fluorescence \cite{tiwari_optical_2020}, autofluorescence \cite{pavlova_understanding_2008}, and optical coherence tomography \cite{badhey_intraoperative_2023} have shown promise for tissue classification in oral lesions, but they require expensive equipment and expert interpretation, which restrict adoption in many primary care and dental environments.

Electrical impedance spectroscopy (EIS) is a non-invasive, non‑ionizing technique that can reveal changes in tissue architecture and cell morphology associated with disease progression.
EIS has been shown to discriminate healthy from cancerous tissue in multiple anatomical sites, including breast \cite{hope_technology_2003}, prostate \cite{halter_electrical_2007}, and the oral cavity\cite{murdoch2014use,lloyd_vivo_2025}. 
The oral cavity is particularly well-suited for EIS because lesions are readily accessible without surgical access. 
Nevertheless, clinical adoption of EIS for oral lesions has been limited. 
Two key barriers are (1) reliably distinguishing OSCC and high risk dysplastic lesions that require intervention from low risk lesions suitable for monitoring, and (2) providing dental clinicians with easy-to-interpret, actionable outputs so they can confidently refer or biopsy when indicated.

Prior EIS investigations of oral lesions have reported promising classification performance, though few in vivo studies have been performed.
Murdoch et al. \cite{murdoch2014use} included \num{51} control subjects and \num{11} patients with oral potentially malignant disorder (OPMD) or oral cancer. 
They reported an area under the curve (AUC) of \num{0.78} for discriminating between in vivo high-risk lesions (OSCC and high-grade dysplasia) and low-risk lesions (low-grade dysplasia and healthy tissue). 
Sun, et. al. \cite{sun_use_2010} and Ching, et. al. \cite{ching_preliminary_2010} explored the use of a 4-electrode bioimpedance probe in the context of tongue cancer detection with \num{12} tongue cancer and healthy subjects at six different frequencies. 
Those studies indicate potential for the use of EIS but are limited by relatively small numbers of OPMD samples.

Recent work with this device using a larger electrode-array EIS device recorded in vivo measurements from 25 subjects and reported high performance (AUC=0.88) for discriminating malignant from healthy tissue, motivating further study with broader patient populations and tissue types \cite{lloyd_vivo_2025}. 
While conventional EIS devices typically inject current at one electrode pair and measure voltage at another, our multi-electrode architecture enables simultaneous voltage measurements across numerous electrode configurations.
This approach, inspired by electrical impedance tomography (EIT) system design but applied to EIS measurement, generates high-dimensional datasets that contain rich diagnostic information but also present analytical challenges distinct from traditional single-pair EIS. 
Conventional threshold-based analyses typically use only one impedance parameter at a time, measured at one frequency; thus, they are constrained in their ability to exploit the complex, multi-electrode, multi-frequency data from our device.

Machine learning (ML) methods have become widely used in medical diagnostics for classification, detection, and segmentation. 
In oral cancer research, several studies have applied ML, particularly convolutional neural networks (CNNs), to a range of imaging modalities and reported encouraging performance. 
Examples include CNN-based classification of biopsy-derived epithelial layer images with test accuracy near \SI{89}{\percent} \cite{gupta2019tissue}, automated lesion classification from the spectral-spacial features of hyperspectral-image lesion imaging \cite{jeyaraj2019computer}, deep-learning combined with traditional classifiers on confocal laser endomicroscopy achieving accuracies around \SI{88}{\percent} \cite{aubreville2017automatic}, and 3D CNNs that improve CT-based diagnosis by capturing inter-slice continuity \cite{Xu2019}. 

However, relatively few studies have applied ML directly to EIS data for oral lesions. 
A recent study by Lin et. al. \cite{lin_deep_2025} applied a CNN directly to EIS data collected from the ventral tongue and floor of mouth (\num{51} healthy controls and \num{11} patients with OPMD or oral cancer), reporting strong discrimination with an AUC of \num{0.907} for classifying high versus (vs) low risk lesions. 
Veil et. al. \cite{veil_electrical_2025} investigated a patient-based classification approach using a Gaussian process classifier (GPC) to differentiate healthy from tumorous bladder tissue using EIS data collected from ex vivo samples of six patients.
Additionally, their team explored methods to reduce the dimensionality of the feature space to improve model efficiency. 
However, the limited number of patients with cancer and OPMD restricts the robustness and generalizability of these models.
Though these models show promising results, the current study does not include CNN or GPC models due to the high risk of overfitting when trained on data where the number of features is less than the number of independent samples.

In this study, we evaluate ML models for three different classification tasks to analyze EIS signals.
ML is used both to maximize diagnostic performance and to identify which subsets of measurements carry the most useful information so that dimensionality can be reduced and classification performed in real time for clinical use. 
We systematically optimize the selection of injection/measurement electrode pairings, measurement frequencies, and signal normalization strategies to (a) reduce computational overhead, (b) improve the reliability and generalization of ML classifiers, and (c) produce outputs appropriate for non-specialist clinicians. 
Our aim is to apply ML to raw EIS signals to improve discrimination among normal mucosa, benign lesions, dysplasia, and OSCC and to provide accurate tissue assessment in real time to aid early diagnostic clinical decisions.

The full code for tuning, evaluation, and results generation is available on GitHub \cite{Thones_IIVV_Evaluation_2025}. 
The data


\section{Methods}
\label{Methods}

\subsection{EIS data acquisition system and probe}

EIS is used to characterize the frequency-dependent electrical properties of a material or biological tissue.
A small alternating current is applied between a pair of electrodes over a range of frequencies while the induced voltage in the object of interest is measured. 
The resulting bioimpedance,
which provides information about the resistive and capacitive properties, can be calculated by a generalized form of Ohm's law to account for the complex impedance resulting from AC current.
This study uses the impedance magnitude $|Z|$ as the primary feature. It consists of a resistive ($R$) and an imaginary reactive component ($X$) and is defined as such,
\begin{equation} 
     |Z| = \sqrt{R^2+X^2}.
     \label{eq:zabs}
\end{equation}

Previously, a custom electrical impedance probe was developed for intraoperative in vivo data collection \cite{lloyd_vivo_2025}. 
The probe interfaces to a \num{32}-channel commercial impedance measurement system (EIT32, Sciospec Scientific Instruments GmbH, Germany), configured to deliver a sinusoidal current at \SI{0.1}{\milli\ampere} peak amplitude over a broad frequency range and record voltages at all available electrode channels.
The probe incorporates an \SI{11}{\milli\metre} diameter electrode array housed in a biocompatible, resin-printed casing (Form 3, Formlabs, USA).
Cables connect the probe to the EIT system so that the probe could remain in the sterile surgical field while the measurement hardware remains outside.

\begin{figure}[t]
    \centering
    \begin{tikzpicture}
        \node[anchor=south west, inner sep=0] (image) at (0,0) {\includegraphics[width=0.6\columnwidth]{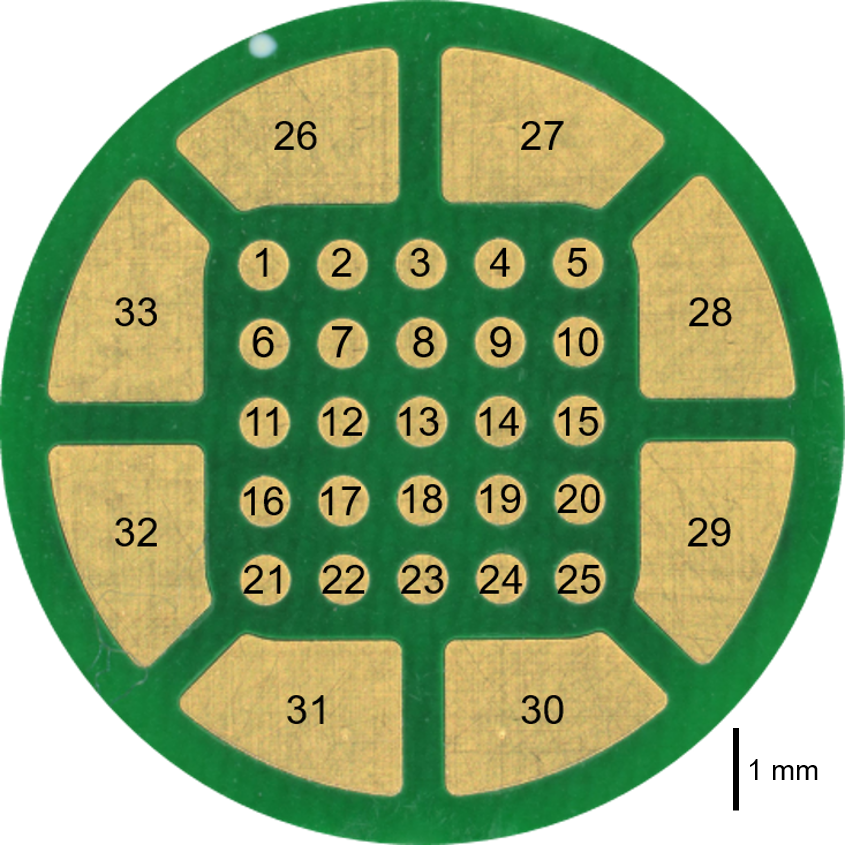}};
        \begin{scope}[x={(image.south east)}, y={(image.north west)}]
        \end{scope}
    \end{tikzpicture}
    \caption{PCB electrode array with numbered electrodes. 
    The center contains a five-by-five array of \SI{0.6}{\milli\metre} electrodes (voltage pick-up electrodes), surrounded by eight larger, current injection electrodes. Electrode 13 was not used due to hardware limitations, resulting in 24 available voltage electrodes.}
    \label{fig:Electrode_array_numbering}
\end{figure}

The measurement array, adapted from previously validated configurations \cite{Kossmann2024, Doussan2025}, was designed for tetrapolar impedance measurements.
The array consists of a $5\times5$ grid of \SI{0.6}{\milli\metre} voltage‑sensing electrodes surrounded by eight larger current‑injection electrodes; the electrode geometry and numbering are shown in Fig.~\ref{fig:Electrode_array_numbering}.
Electrode \num{13} is excluded from measurements because of the \num{32}‑channel limitation of the measurement system, thus only \num{24} voltage measurement electrodes were used.

Tetrapolar drive-sense patterns, using spatially separated current drive and sink electrodes (II) and voltage electrode pairs (VV), are employed to reduce capacitive effects at the electrode–tissue interface, which are particularly relevant at low frequencies. 
Larger peripheral electrodes, used for current injection, provide greater surface area, reduce contact impedance, and improve the signal-to-noise ratio \cite{Kossmann2024}.
High‑input‑impedance amplifiers on the voltage channels (small $5\times5$ grid of electrodes) minimize current draw and associated measurement artifacts. 
The electrode configuration supports \num{28} unique II combinations and \num{276} distinct VV pairs among the \num{24} active inner electrodes, yielding \num{7728} unique current‑injection‑voltage‑measurement (IIVV) permutations. 
Impedance spectra were acquired for all IIVV combinations at \num{31} logarithmically spaced frequencies from \SI{100}{\hertz} to \SI{100}{\kilo\hertz}.
All 28 current injection patterns and simultaneous voltage electrode measurements for each full spectral frame takes approximately \SI{3}{\second} to be recorded. 

\subsection{Clinical study protocol}
In total, \num{114} adult patients undergoing resection or surgical biopsy of oral cavity or oral pharynx squamous cell carcinoma were enrolled in this study after providing written informed consent (NCT05430477). 
The study protocol was approved by the Institutional Review Board at Dartmouth-Hitchcock Medical Center (IRB\#02001317). 
Patients were treated according to institutional standard of care for surgical treatment and specimen histopathological assessment. 
Low temperature hydrogen peroxide gas plasma (STERRAD) treatment was utilized to sterilize impedance devices prior to in vivo usage. 
The primary intraoperative in vivo EIS measurements were obtained from the center of the lesion, with additional recordings acquired from a healthy contralateral site when available.
In some cases, serial measurements were taken beginning at the lesion center and extending toward the margin or contralateral side to capture regions of dysplasia and non-malignant tissue, yielding up to five recording sites across the lesion and surrounding tissue. 
At each site, the probe face was placed in contact with the oral mucosa and held stationary for approximately \SI{10}{\second} while three consecutive impedance spectra were acquired.
The surgeon then repositioned the probe at the same location, and this process was repeated twice more to generate repeated measurements, resulting in a total of nine spectra per site. 
Measurement sites were marked on the tissue in vivo with a surgical pen, and ink was subsequently applied to the probe contact area on the ex vivo specimen to enable pathological correlation.

To align our analysis with clinical decision-making, probed tissue samples were grouped into four clinically relevant categories: (1) healthy tissue, (2) confirmed OSCC, (3) high-grade dysplasia and carcinoma in situ (CIS), and (4) non-malignant lesions, which includes low-moderate grade dysplasia, hyperkeratosis, and normal/benign tissue collected from lesion or near-lesion sites.

To address the well‑documented variability in dysplasia grading, we adopted a two‑tier system that collapses mild and mild‑moderate dysplasia into a low‑grade category and combines moderate, severe dysplasia, and CIS into a high‑grade category \cite{kujan_why_2007,speight_interobserver_2015}. 
This binary grouping, which has been shown to enhance diagnostic consistency and better predict malignant transformation risk compared with the conventional three‑tier scheme \cite{ellonen2025binary, mannapperuma_oral_2025}, was used for all subsequent analyses.

\subsection{Data pre-processing}
Each EIS sample yields a complex-valued impedance dataset of fixed dimensions $\num{7728} \times \num{31} \times \num{9}$, where $N_{IIVV}=7728$ corresponds to unique measurement geometries, $N_{freq}=31$ represents logarithmically-spaced frequencies from \SI{100}{\hertz} to \SI{100}{\kilo\hertz}, and \num{9} repeats consists of three repeated test measurements, each containing three consecutive burst measurements. 

Individual burst measurements were processed using a multi-step filtering procedure adapted from \cite{lloyd_vivo_2025}. 
First, voltage-based filtering identifies and excludes poor-quality signals. 
IIVV patterns corresponding to these excluded bursts are then removed.
A subsequent filtering step applies impedance-based criteria to eliminate additional artifacts. 
To maximize data quality and retain as many valid IIVV patterns per tissue sample as possible, measurements are averaged within bursts and then across repeated tests, yielding a $\num{7728} \times \num{31}$ dataset with some number of IIVV patterns removed. 

To control data quality, a completeness threshold ($C_{th}$) was set as the maximum allowable percentage of these removed IIVV patterns per sample. 
Only samples meeting this completeness criterion $C_{th}$ = \num{60}\% were included for further analysis.

To characterize input size, we define the scalar $N_{input}$ as the product of the number of frequency bins and the number of IIVV patterns: 
\begin{equation}
    N_{input} = N_{freq} \cdot N_{IIVV}.
\end{equation}
This metric quantifies the dimensionality of the EIS data fed into the classification model and reflects the computational complexity across experimental configurations.

Finally, z‑score normalization was applied to all input data per-frequency.
Because tissue impedance can differ by an order of magnitude across the frequency spectrum, normalizing each frequency separately prevents the model from being biased toward high‑magnitude features.
This normalization accelerates training, improves stability, and enhances overall predictive accuracy.

\subsection{Optimization schedule}
\label{subsec:opt_schedule}

\begin{figure}[ht]
    \centering
    \includegraphics[width=1\columnwidth]{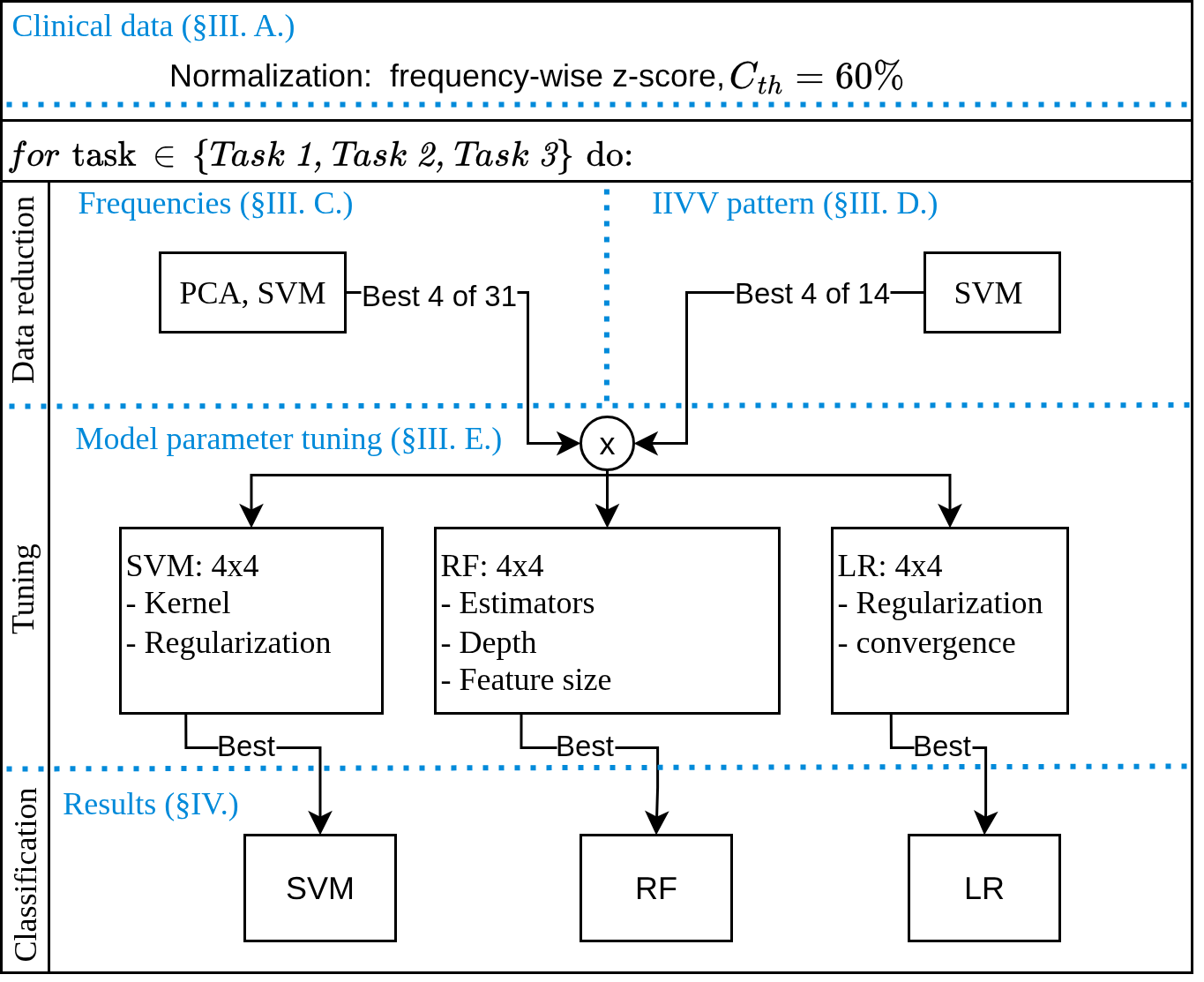}
    \caption{Optimization schedule for reducing the amount of EIS data sample components needed for a reliable classification. Details for each step in the optimization schedule are described in the manuscript sections referenced parenthetically.}
    \label{fig:Optimization_schedule}
\end{figure}

Given the high dimensionality of the dataset, we aimed to systematically reduce the number of classification input parameters to minimize computational load while improving clinical efficiency and decision reliability. 
The evaluation and optimization workflow is illustrated in Fig.~\ref{fig:Optimization_schedule}.

Let $\mathbf{x}_n$ denote the impedance magnitude feature vector of the \textit{n-th} data sample, computed by (\ref{eq:zabs}).
Each $\mathbf{x}_n$ comprises $N_{input}$ features, where the number of features is determined by the set of IIVV patterns and the selected frequencies. 
When \textit{All} available IIVV patterns and frequencies are employed, the input dimension becomes $N_{input}=\num{239537}$ features per sample, which yields approximately \num{58.4} million inputs across the full EIS dataset. 

For a given classification task the number of samples varies with the tissue type and class labels; we denote this quantity by $N_{task}$.
Accordingly, each task is represented by a dataset
\begin{equation}
\begin{aligned}
\{\,\mathbf{X},\mathbf{Y}\,\}_{n=1}^{N_{\mathrm{task}}} &
    \begin{cases}
        \mathbf{X} = \{\mathbf{x}_{1},\mathbf{x}_{2},\dots,
                \mathbf{x}_{N_{\mathrm{task}}}\}\\
          \mathbf{Y} = \{\mathbf{y}_{1},\mathbf{y}_{2},\dots,
                \mathbf{y}_{N_{\mathrm{task}}}\},
    \end{cases} 
\end{aligned}
\end{equation}

where $\mathbf{X}$ contains the prepared EIS measurement vectors and $\mathbf{Y}$ holds the corresponding class labels. 

As a baseline, classification results were obtained using all frequencies and IIVV patterns for each task. 
All subsequent dimensionality reductions and tuning steps were evaluated relative to this baseline. 
Model performance evaluation throughout the IIVV pattern and frequency optimization was done using the default support vector machine (SVM) implementation from scikit-learn (version 1.7.2, \cite{Scikit-learn}), chosen as a heuristically determined baseline model.


The top four parameter sets identified from frequency reduction (detailed in \ref{sub:Frequency_reduction}) and the IIVV pattern reduction in \ref{sub:IIVV_pattern_reduction} were combined in a multi-stage model tuning process outlined in \ref{sub:Model_and_parameter_tuning}.
All \num{16} possible combinations (four frequency sets times four IIVV sets) were evaluated using SVM, Random Forest (RF), and Logistic Regression (LR) classifiers.
For each combination, a subsequent model-specific parameter tuning was conducted as described in \ref{sub:Model_and_parameter_tuning}. 

The best-performing parameter set and model were selected individually for each classification task (\ref{sub:Model_and_parameter_tuning}). 
Accuracy, F1-score, and AUC were used as the scoring metrics for feature optimization.
Final performance metrics for all models and classification tasks are presented in the results Section \ref{sec:Results}.

\subsubsection{Classification tasks}
\label{subsub:class_task}

Three different classification tasks were evaluated.
\textit{Task 1} is a binary classification distinguishing healthy from cancer samples.
\textit{Task 2} is a multi-class classification of the lesion pathologies: cancer, high-grade dysplasia, and non-malignant.
\textit{Task 3} is a multi-class classification separating four categories: healthy, cancer, high-grade dysplasia, and non-malignant lesions.

\subsubsection{Reducing top frequencies}
\label{sub:Reducing_top_frequencies}
Principal component analysis (PCA) was used to identify the most informative frequencies for classification. 
PCA is a standard technique for dimensionality reduction that transforms correlated input variables into a smaller set of orthogonal principal components that capture the dominant variance in the data \cite{greenacre2022principal}.
We hypothesize that the full measured impedance data space is redundantly high-dimensional for the classification task and that a small number of principal components would suffice to separate classes.
PCA was therefore applied to the EIS spectra to quantify each frequency’s contribution to the components that drive class separation; frequencies with low contributions were treated as less informative and candidates for exclusion.

We evaluated frequency importance using two complementary objectives: (1) rank frequencies by their information content and (2) determine the minimum number of frequencies required to achieve satisfactory classification performance. 
Patient-level five‑fold cross-validation was used to generate independent train/test splits. 
Within each fold, PCA was applied to the training data to derive a frequency ranking; the ranked frequencies were then used sequentially as inputs to the default scikit-learn SVM classifier to quantify classification performance as a function of the number of frequencies included.

\subsubsection{Reducing IIVV patterns}
\label{sub:Reducing_IIVV_pattern}
Due to the electrode-array configuration and the exhaustive combinations of current injection (II) and voltage measurement (VV), many IIVV sets yield negligible voltage differences. 
This occurs when voltage electrodes are aligned approximately parallel to equipotential regions, are positioned very close together, or are distant from the current-injection pair.
We hypothesized that not all IIVV patterns provide meaningful information about tissue electrical properties; some may introduce noise or reduce impedance contrast.

To evaluate this, we systematically assessed \num{13} sets of geometry-based IIVV patterns by applying masks to specify which patterns to include in a set, allowing controlled analysis of different pattern subsets. 
The geometric groups were based on four different patterns of current injection electrode pair geometries: \textit{adjacent}, \textit{skip 1}, \textit{opposite}, and 
\textit{across}. 
Within the \textit{adjacent}, \textit{skip 1}, and \textit{opposite} groupings, the rows or columns of the voltage pair electrodes are described as either far, medium, or close. 
For the far sets, all voltage electrodes in the two rows/columns furthest from the current injection electrode pair are included (eg. for \textit{Adj. far} II: 26-27, rows with voltage electrodes 16 and 21).
All combination of electrodes in those designated rows/columns are used in the IIVV pattern set. 
For the across (+) patterns (eg. \textit{Long/med. a+}), only current electrode pairs on opposite sides of the voltage electrode array are used, with electrodes in the rows/columns aligning with that side as the VVs. 
The \textit{Long a+} set only includes the 2 voltage electrodes in the same rows or columns as the current drive electrodes, while the extended versions of the patterns sets include all voltage electrodes in the rows. 
Table~\ref{tab:masking_overview} lists all tested masks and their corresponding impact on the IIVV input dimension $N_{IIVV}$ (see GitHub repository \cite{Thones_IIVV_Evaluation_2025} for IIVV compositions images). 
The \textit{All} mask includes every IIVV pattern and serves as the baseline for comparison.

Classification performance was also evaluated using a diverse IIVV pattern set containing multiple geometries. 
This set was created by thresholding the averages of all measured impedance values at \SI{100}{\hertz} into two groups: those with average impedance values below the threshold and those with values above it. 
The \textit{z-threshold} was varied across the full range of impedance values and the resulting set of IIVV patterns were then applied to each individual sample to test the classification performance. 

\begin{table}[ht]
\centering
\caption{IIVV masking overview for the 13 geometry-based IIVV pattern with the number of parameters and the number of II and VV electrodes used, along with average II and VV distances in mm.}
    \begin{tabular}{
        @{} l 
        S[table-format=4]
        S[table-format=1]
        S[table-format=2]
        S[table-format=2.2]
        S[table-format=2.2] @{}}
\toprule
{IIVV mask} & {$N_{IIVV}$} & {II} & {VV} & {II dist.} & {VV dist.} \\
\midrule
            \textit{All}           & 7728 & 8 & 24 & 4.54 & 2.71 \\
            \textit{Long a+}       &   16 & 8 & 12 & 6.16 & 4.06 \\
            \textit{Long a+ ext.}  &  120 & 8 & 16 & 6.16 & 3.16 \\
            \textit{Med. a+ ext.}  &  120 & 8 & 16 & 6.16 & 2.06 \\
            \textit{Skip1 close}   &  360 & 8 & 24 & 4.36 & 2.01 \\
            \textit{Adj. close}    &  292 & 8 & 24 & 2.83 & 1.80 \\
            \textit{Med. adj.}     &  387 & 8 & 20 & 2.83 & 2.15 \\
            \textit{Adj. far}      &  660 & 8 & 24 & 2.83 & 2.28 \\
            \textit{Skip1 medium}  &  440 & 8 & 20 & 4.36 & 2.16 \\
            \textit{Skip1 far}     &  224 & 8 & 24 & 4.36 & 1.58 \\
            \textit{Opp. close}    &  264 & 8 & 24 & 6.16 & 2.48 \\
            \textit{Opp. medium}   &  264 & 8 & 24 & 6.16 & 2.48 \\
            \textit{Opp. far}      &  264 & 8 & 24 & 6.16 & 2.69 \\
        \bottomrule
\end{tabular}
\label{tab:masking_overview}
\end{table}

For each classification task, data filtered by each mask were input into the default SVM model. 
To ensure robustness, we performed five randomly initialized trials with a five-fold patient group-wise train/test split. 
The input and output data structure is described in \ref{sub:Cross-val}. 
For all IIVV masks, the full set of \num{31} frequency measurements across the EIS spectrum was used. 

\subsubsection{Models}
SVM, RF, and LR classifiers were trained and tested on the defined classification tasks.
A practical advantage of these models is their ability to accommodate variable input dimensionality without requiring architectural changes, facilitating straightforward retraining with reduced feature sets.

The SVM classifier aims to find an optimal separating hyperplane by implicitly mapping non-linear input data into a high-dimensional feature space using kernel functions \cite{cortes1995support, Hearst1998}. 
Different kernels create linear and non-linear decision boundaries, as they compute similarities between samples in a transformed feature space without explicitly mapping all data points. This allows SVMs to efficiently capture complex, non-linear relationships.

The RF model is an ensemble of decision trees, each trained on different random subsets of features and samples drawn with replacement \cite{breiman2001random}.
This randomness promotes decorrelation among trees, enhancing generalization.

LR models estimate class probabilities by applying a logistic sigmoid function to a weighted linear combination of input features \cite{david2000applied}.
Regularization is applied to prevent overfitting and improve generalization.

Hyperparameter tuning for all models is detailed in 
\ref{sub:Model_and_parameter_tuning}.
For SVM, both kernel function and regularization parameter \textit{C} were optimized. 
For RF, tuning involved estimators, tree depth, and feature subset size. 
For LR, the regularization strength and solver convergence parameters were adjusted to ensure stable, generalized solutions.

Furthermore, all hyperparameter tuning procedures were conducted with respect to the optimization targets: F1-score, accuracy, and AUC, as described in the following section. The complete tuning methodology and corresponding results are documented in \cite{Thones_IIVV_Evaluation_2025}. 
Results obtained using AUC as the target metric are reported, as this criterion yielded the most robust performance across all evaluated tasks.

\subsubsection{Evaluation Metrics}
To evaluate the impact of the parameter optimization, precision, recall, F1-Score, accuracy, and the Receiver Operator Characteristic curve (ROC) AUC metrics were used in this study. 

Precision gives the ratio of true positives among the samples predicted as positive by the model. It is defined as
\begin{equation}
    P = \frac{TP}{TP + FP},
    \label{eq:Precision}
\end{equation}
where $TP$ is the number of true positive samples and $FP$ is the number of false positive samples. 

Recall measures the proportion of actual positive instances that are correctly identified by a model. 
It is defined as 
\begin{equation}
    R = \frac{TP}{TP + FN},
    \label{eq:Recall}
\end{equation}
where $FN$ is the number of false negative predictions. Recall is an important metric to use in situations where a false negative is more problematic than a false positive, such as here, where undetected cancer leads to later stage treatment with worse outcomes. 

The F1-score is often used in ML approaches as a robust method to evaluate model performance with unbalanced data. 
It is the harmonic mean of precision and recall, defined by 

\begin{equation}
    \text{F1} = \frac{2\,TP}{2\,TP + FP + FN}.
    \label{eq:f1}
\end{equation}

The classification accuracy is defined as the proportion of correctly predicted instances among the total number of instances.
 
ROC curves plot the $TP$ rate against the $FP$ rate across all classification thresholds. 
The AUC summarizes classifier performance, where a value of 1 indicates perfect discrimination, and 0.5 represents random guessing of binary classes \cite{hanley1982meaning}. 
For each feature reduction step, the best parameter set was determined using the model AUCs. 
For the multiclass classification problems, we report the micro‑average ROC curve and its corresponding micro‑average AUC. Micro‑averaging takes into account class imbalances and aggregates the decisions over classes by summing the individual $TP$ and $FP$ counts before computing the rates; thus the resulting curve reflects the classifier’s overall ability to distinguish any positive instance from any negative one, regardless of class label.
Micro-averaged scores are reported for multiclass tasks during baseline and final results creation. Micro-averaging takes into account sample imbalances between the individual tasks.
During the tuning procedures macro-averaging was used to optimize the models in terms of the target classes, weighing the performance of each class equally to encourage classification of the lower-represented tissue types. 

Paired t-tests were performed between the AUCs of the baseline model and candidate models for the selected parameters from each training fold, for each task separately.
Statistical significance is determined with an alpha level of 0.05. 

\section{Model training and tuning}
\label{sec:Model_training_and_tuning}

\subsection{Clinical data}
\label{sub:Clinical_data}
\begin{figure}[ht]
    \centering
    \begin{tikzpicture}
        \node[anchor=south west, inner sep=0] (image) at (0,0) {\includegraphics[width=\columnwidth]{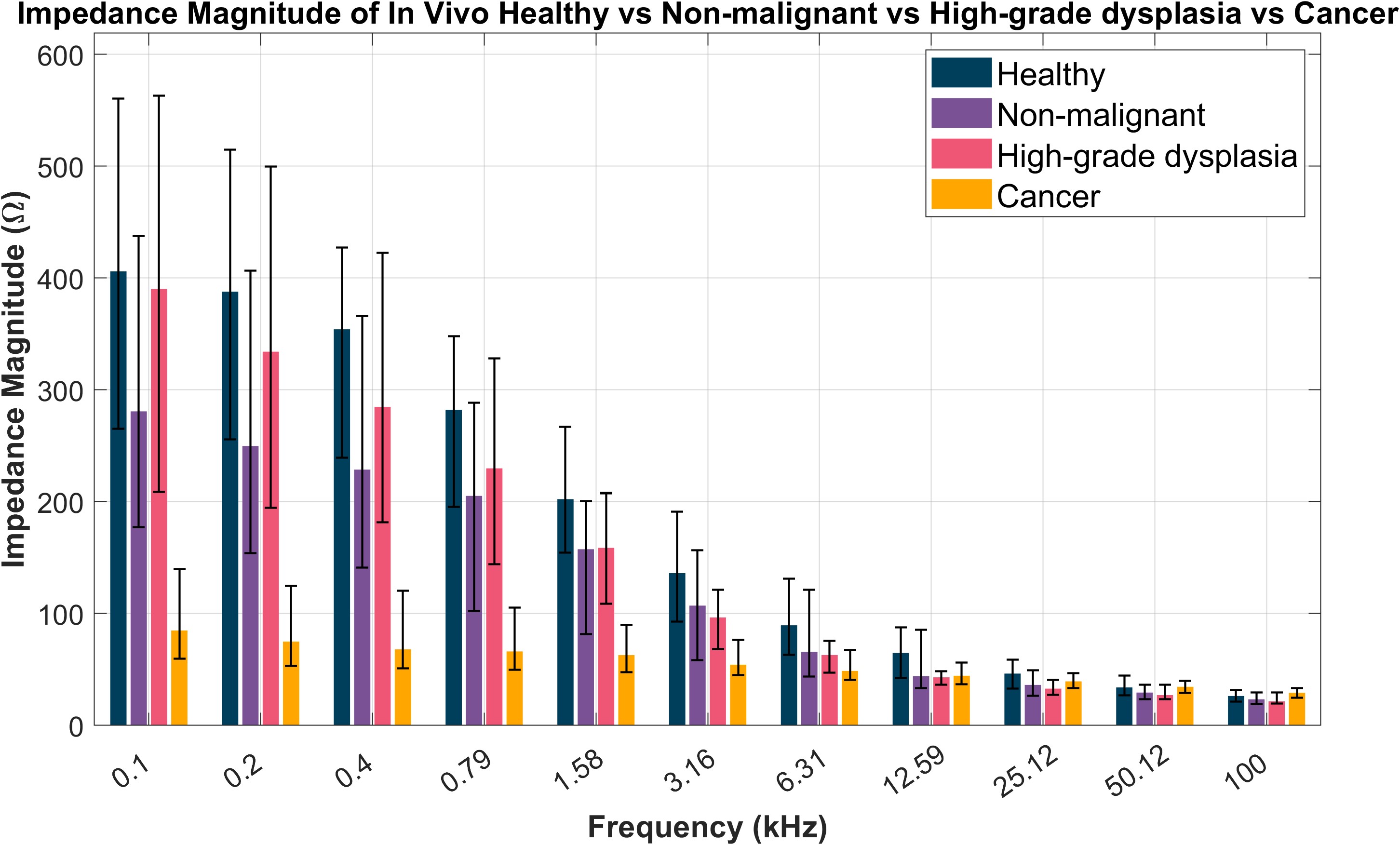}};
        \begin{scope}[x={(image.south east)}, y={(image.north west)}]
        \end{scope}
    \end{tikzpicture}
    \caption{The impedance magnitude of all four tissue types across frequencies.
    The average of the impedance magnitude of the adjacent far IIVV pattern set was used to compute the impedance spectra for each sample. Error bars show the standard deviation on each tissue class.}
    \label{fig:ZMag-tissues}
\end{figure}

Valid data were collected from \num{104} subjects, with multiple measurements from each subject. 
In total, electrical impedance was measured at \num{276} independent sample sites, with \num{100} healthy tissue samples, \num{96} OSCC lesions, \num{44} non-malignant lesions, \num{32} high-grade dysplasia or carcinoma in situ samples, and \num{4} non-dysplastic or OSCC pathologies that were excluded due to low sample numbers.
See Fig.~\ref{fig:ZMag-tissues} for the average impedance magnitude across all tissue types.

An extensive evaluation of various combinations of completeness thresholds and normalization methods indicated that applying z-score normalization along the frequency axis, combined with a completeness threshold of $C_{th}=\,$~\SI{60}{\percent}, yielded optimal classification performance.
This $C_{th}$ yielded \num{256} viable samples to be used in data analysis (see Table~\ref{tab:demographics}). 
The completeness threshold and normalization method were fixed throughout this study to reduce the complexity of performance evaluations.

\begin{table}[ht]
  \caption{Subject and sample pathology of the dataset with the data available for different completeness threshold ($C_{th}$).}
  \label{tab:demographics}%
  \centering
  \begin{tabular}{@{}l S[table-format=2] S[table-format=2] S[table-format=2] S[table-format=2] @{}}
    \toprule
     & \multicolumn{2}{c}{Number of subjects}
     & \multicolumn{2}{c}{Number of samples $N$} \\
     \cmidrule(lr){2-3} \cmidrule(lr){4-5}
     Type      
      & {$C_{th}$=\,\SI{100}{\percent}}
      & {$C_{th}$=\,\SI{60}{\percent}}
      & {$C_{th}$=\,\SI{100}{\percent}}
      & {$C_{th}$=\,\SI{60}{\percent}} \\
    \midrule
    Healthy              & 95 & 88 & 100 & 90 \\
    Cancer               & 70 & 70 & 96 & 96 \\
    High-grade dysp.     & 21 & 19 & 32 & 29 \\
    Non-malignant        & 28 & 27 & 44 & 41 \\
    Other                &  4 &  4 &   4 &  4 \\
    \bottomrule
  \end{tabular}
\end{table}

\subsection{Cross-validation}
\label{sub:Cross-val}

To assess the generalization capability of the models across different patient groups, a leave-one-patient-group-out cross-validation (LOPGO-CV) was employed. 
During each LOPGO-CV iteration, one entire patient group is withheld as the test set, while all remaining patients are used for training. 
This ensures that the model is evaluated exclusively on patient data it has never encountered during training, thereby providing a measure of unknown patient-level robustness.
A five-fold LOPGO-CV is defined by dividing all available patients into five groups of approximately equal size. 
Each group is used once as the test set, while the remaining four groups form the training set. 
Performance metrics are aggregated from all folds to obtain an unbiased estimate of the model’s ability to generalize on unseen patients.


\subsection{Frequency reduction}
\label{sub:Frequency_reduction}

Frequency reduction tasks were first performed on \textit{All} training data, to be independent of IIVV pattern selection. 
PCA was performed on the training data, and loadings from the first \num{10} principal components were extracted to capture \SI{99}{\percent} of the variation in the dataset. 
A frequency importance score was calculated for each frequency by summing the squared loadings across the \num{10} components. 
Frequencies were then ranked in descending order by this score, with higher loadings indicating greater importance. 
Rankings were determined within each cross-validation fold and aggregated across folds to generate a composite ranking.

Using the PCA-derived frequency ordering from each fold, reduced datasets were created by selecting the highest-scoring $\fT$ frequencies, with $\mathbf{f}_{\mathrm{T} = 1}$ representing the single most informative frequency and $\mathbf{f}_{\mathrm{T} = 31}$ including all frequencies ranked by importance. 
The number of selected frequencies, $N_{\text{freq}}$, varied according to the length of $\fT$. 
Each reduced training set was flattened and used as input to an SVM classifier, which was trained on the training patients and evaluated on the held-out test group. 
Probabilistic outputs from the SVM were used to train the model for the three optimization scores: AUC, accuracy (Acc), and F1.


\subsubsection{Task 1}
Fold-wise AUCs and the mean AUC across folds were recorded for each number of top frequencies $\fT$, generating a performance curve of mean AUC vs the number of included frequencies.
The fold-wise PCA rankings were also converted into a point‑scoring scheme and aggregated across folds to produce the composite frequency ranking. 

\begin{figure}[th]
    \centering
    \includegraphics[width=0.9\columnwidth]{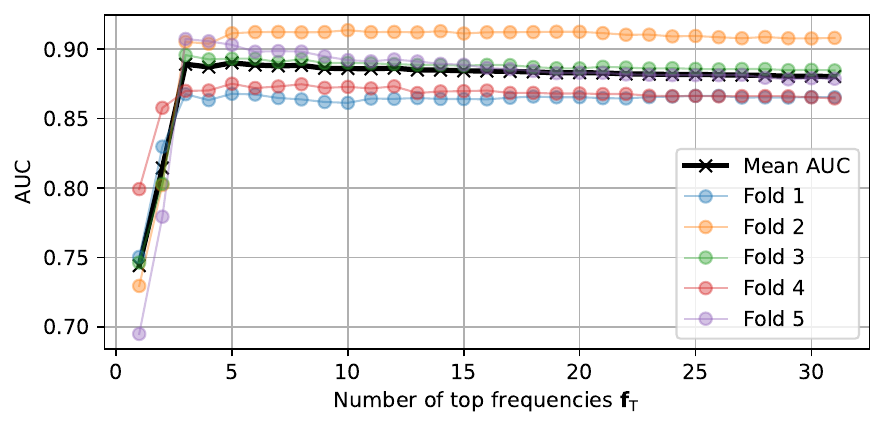}
    \caption{\textit{Task 1} classification AUCs using all IIVVs and reduced frequencies, with AUC as the training metric. Each line shows the averaged AUC from ten trials of five-fold cross-validation.}
    \label{fig:baselineaucs_healthy_vs_cancer}
\end{figure}

Results from each fold of the first of ten trials are shown in Fig.~\ref{fig:baselineaucs_healthy_vs_cancer}.
The AUC increased sharply with the first few top frequencies and plateaued after approximately the top four frequencies. The modest variability observed among folds likely reflects differences in feature representation across patient splits. Fig.~\ref{fig:freq_reduction_all_tasks} plots the \SI{95}{\percent} confidence interval of AUCs from all ten trials. This implies that \textit{Task 1} benefited from using fewer frequencies.

\subsubsection{Task 2}
The average AUCs for \textit{Task 2} were lower than those of \textit{Task 1} for all frequency groups, with a baseline AUC of \num{0.696} (Fig.~\ref{fig:freq_reduction_all_tasks}). 
No reduced frequency dataset significantly improved the baseline AUC, (i.e., p-values $>\num{0.05}$). 
The reduced datasets from  $\mathbf{f}_{\mathrm{T} = 15}$, $\mathbf{f}_{\mathrm{T} = 23}$, and $\mathbf{f}_{\mathrm{T} = 2}$ all resulted in AUCs of 0.70, which non-significantly improved classification performance as compared to the baseline result.


\subsubsection{Task 3}
\begin{figure}[ht]
    \centering
    \includegraphics[width=0.9\columnwidth]{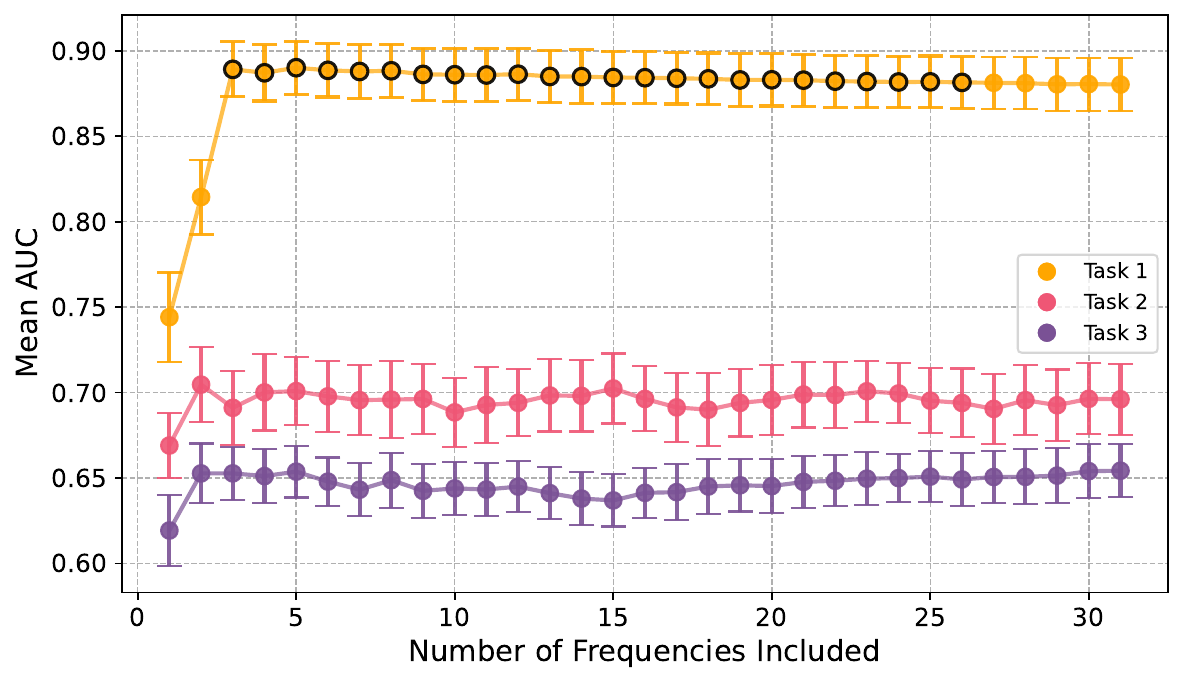}
    \caption{Reduced Frequency \SI{95}{\percent} Confidence Interval of Mean AUCs for All Tasks. CI shown by error bar, black outline on colored markers designate statistically higher AUC as compared to all frequencies (p $<\num{0.05}$).}
    \label{fig:freq_reduction_all_tasks}
\end{figure}

All reduced frequencies performed equally or worse than the baseline which achieved an average AUC of \num{0.66}, as shown in Fig.~\ref{fig:freq_reduction_all_tasks}.
In the case of multiclass classification, the model may benefit from having as much information as possible. 
Particularly with the high-grade dysplasia and non-malignant classes, there could not be enough data to yield many frequencies redundant, as compared to how few frequencies are needed for the simpler binary classification problems. 
The four top performing frequency groups to be used with the tuned models were: $\mathbf{f}_{\mathrm{T} = 5,30,3,2}$ which had average AUCs of 0.65.

\subsection{IIVV pattern reduction}
\label{sub:IIVV_pattern_reduction}
Classification performance across tasks was evaluated using the same SVM model and scoring metrics described in \ref{sub:Frequency_reduction}. 
Unlike frequency reduction, this phase used two methods to reduce the number of parameters: geometry-based IIVV sets, and impedance threshold (\textit{z-threshold}) set. Both utilize the full frequency spectrum for each evaluation. 
The \textit{z-threshold} that resulted in the highest mean AUC was used in the validation test with all IIVV pattern sets.

\subsubsection{Task 1}
\begin{figure}[ht]
    \centering
    \includegraphics[width=1\columnwidth]{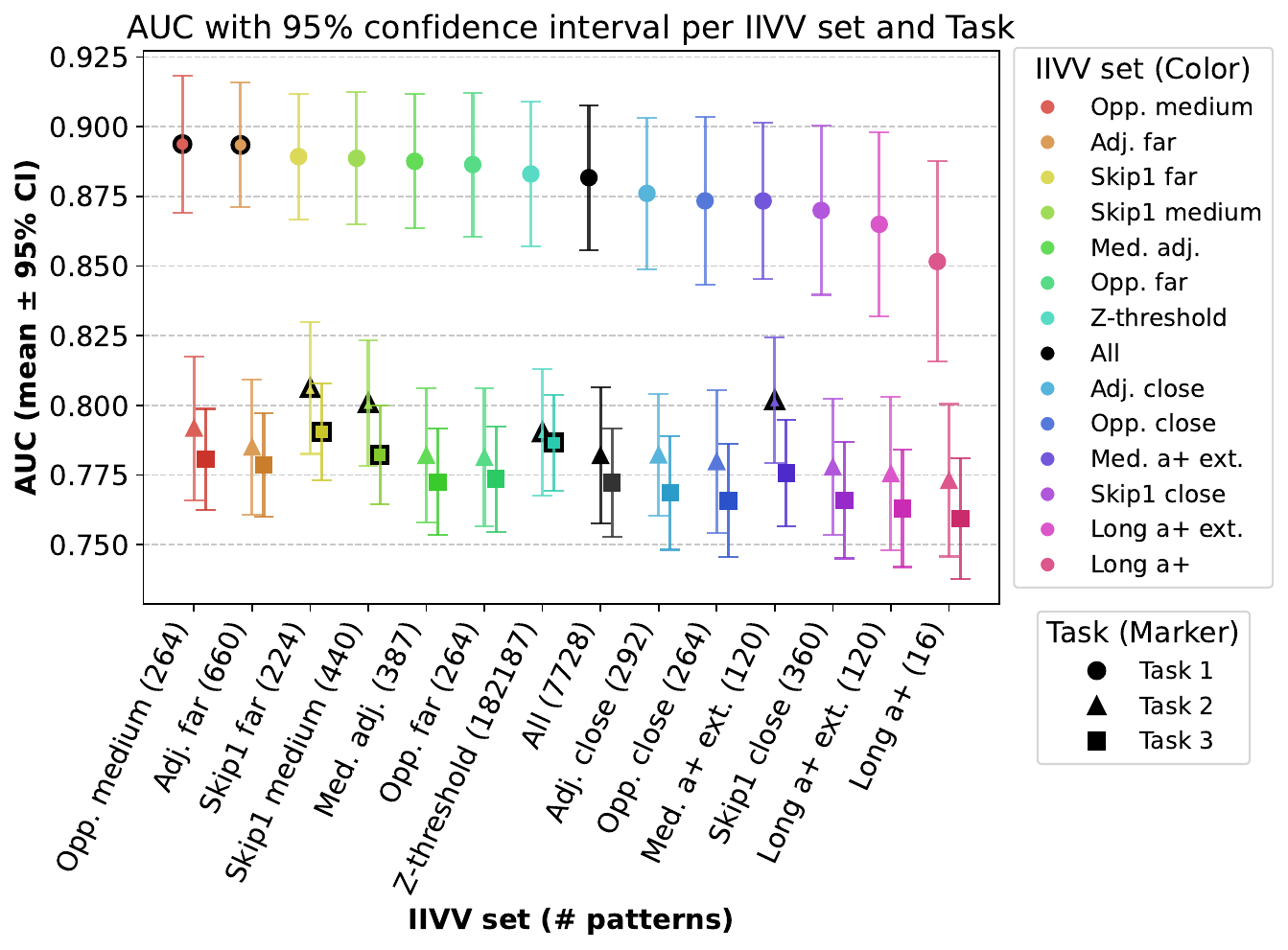}
    \caption{
    IIVV selection results for all tasks, tuned with AUC scoring. 
    The \SI{95}{\percent} confidence intervals of average AUCs IIVV pattern sets are shown with the bars.
    Markers are outlined when the AUC was significantly ($\text{p}< \num{0.05}$) higher than \textit{All} IIVVs (black marker). The number of IIVV patterns in each set is shown parenthetically, for z-threshold, the average number of patterns from all three tasks is shown.}
    \label{fig:IIVV_sel_heatmap_healthy_vs_cancer}
\end{figure}

Fig.~\ref{fig:IIVV_sel_heatmap_healthy_vs_cancer} presents the \SI{95}{\percent} confidence intervals of AUC values from ten five-fold LOPGO-CV trials for each IIVV pattern group, sorted by descending AUCs based on \textit{Task 1}.
The results demonstrate that models trained on selected IIVV subsets maintain strong classification performance (see Table~\ref{tab:masking_overview} for pattern definitions). 
The \textit{All} IIVV pattern, which incorporates all \num{7728} available IIVV patterns and frequency data points, ranked mid-range in performance with an average AUC of \num{0.88}, although it contains substantially more data than the other subsets.
The \textit{Opp. medium} had the highest mean AUC of \num{0.89}. 
The IIVV sets \textit{Adj. far}, \textit{Skip1 far}, and \textit{Skip1. medium} also showed stable AUCs compared to the baseline results. 


\subsubsection{Task 2}
The \textit{Skip1 far} IIVV patterns scored the highest mean AUC of \num{0.81}. 
The next four highest AUC scores were achieved by the IIVV pattern groups \textit{Med. a+ ext.}, \textit{Skip1 medium}, \textit{Opp. medium}, and \textit{z-threshold} of \num{1232} Ohms. 
All of the top five IIVV pattern groups showed statistically significant AUC improvements from the baseline, which scored an average AUC of 0.74 on the cross-validation.
\textit{All} patterns ranked the ninth-best pattern group, indicating that most of the reduced datasets contain the necessary information to make the best classification decision.

\subsubsection{Task 3}
The best-performing IIVV pattern set was \textit{Skip1 far} with a mean AUC of \num{0.79}. 
The z-threshold obtained the second highest mean AUC with the threshold at \num{779} Ohms (including \num{4359} IIVV patterns).
The next best-performing pattern sets were \textit{Skip1 medium}, \textit{Opp. close}, and \textit{Adj. far}.
The \textit{All} pattern was ranked ninth with a mean AUC of \num{0.77}.

These findings reinforce that focused selection of specific IIVV patterns can outperform a model trained on the entire dataset. 

\subsection{Model parameter tuning}
\label{sub:Model_and_parameter_tuning}
For SVM classification, kernel type and the regularization parameter \textit{C} were selected via parameter tuning to balance the trade-off between margin maximization and classification error.
Grid search was used to tune kernel type and regularization strength. 
Tested kernels include polynomial (\textit{poly}), \textit{sigmoid}, radial basis function (\textit{rbf}), and \textit{linear}. 
The regularization parameter \textit{C} was varied from \num{0.001} to \num{1000} in six logarithmic decimal increments. 
This tuning aims to identify model parameters to set up the optimal decision boundary hyperplane for the corresponding classification task \cite{Scikit-learn}.

For the RF model, key hyperparameters tuned by grid search included the number of trees (estimators), maximum tree depth, and the number of features considered for the best split.
The RF model’s generalization error typically stabilizes with increasing trees, providing robustness against overfitting \cite{breiman2001random}.

For the LR classifier, the regularization strength and maximum solver iterations were systematically tuned to optimize performance.
Regularization is applied by default in the implementation used \cite{Scikit-learn}.

Model performance during parameter tuning was assessed with AUC, Accuracy, and F1-score. 
To ensure robustness, scores were averaged over ten trials of randomly initialized five-fold LOPGO-CV employing patient-based training and testing splits. 
Each individual tuning process comprised \num{160} different full five-fold LOPGO-CV comprised of 10 random states from four IIVV masks combined with four different frequency sets.

\section{Results}
\label{sec:Results}
Results from testing the top parameters for frequency reduction and IIVV pattern set selections separately are reported. 
Models were trained and tuned using all combinations of the top four parameters from each reduction technique.
The classification performance from the best performing combinations for each task and tuned model was evaluated with five-fold LOPGO-CV.
Probabilistic class prediction from the unseen patient group of each fold were aggregated to be used for the ROC analysis and classification performance scoring.

\subsection{Frequency reduction}
For all three classification tasks, frequency 31 (\SI{100}{\kilo\hertz}), frequency 1 (\SI{100}{\hertz}) and frequency 13 (\SI{1.58}{\kilo\hertz}) emerged as the three most important frequencies. 
The next most frequently occurring common frequencies across all three tasks were \SI{79.4}{\kilo\hertz} and \SI{630}{\hertz}, indicating that both low and high frequencies contain important information for tissue classification. Table~\ref{tab:Best4Freq_reduction} reports the resulting AUCs from the top frequency groups.

\textit{Task 1} required the least amount of frequencies for classification, with eight or fewer used as input, improving AUCs compared to the baseline of all 31 frequencies. 
Notably, no reduced frequency set in the multiclass \textit{Task 2} or \textit{Task 3} showed an improvement or better results over the baseline AUCs.


Comparing AUCs of baseline SVM models trained on different frequency groups revealed modest differences of only \SIrange{1}{5}{\percent} between the best and worst groups across all tasks, indicating the robustness to the choice of frequency subsets among the selected frequency groups (see Fig~\ref{fig:freq_reduction_all_tasks}).

\begin{table}[ht]
  \centering
  \caption{Best four rankings for the frequency reduction based on the AUC tuning metric for the different tasks using \textit{All} IIVVs.}
  \label{tab:Best4Freq_reduction}
\begin{tabular}{@{}
    S[table-format=1]
    S[table-format=2]
    S[table-format=1.3]
    S[table-format=2]
    S[table-format=1.3]
    S[table-format=2]
    S[table-format=1.3]
  @{}}
    \toprule
         & \multicolumn{2}{c}{\textit{Task 1}}
         & \multicolumn{2}{c}{\textit{Task 2}}
         & \multicolumn{2}{c}{\textit{Task 3}} \\
    \cmidrule(lr){2-3} \cmidrule(lr){4-5} \cmidrule(lr){6-7}
         {Rank} & {$\fT$} & {AUC} & {$\fT$} & {AUC} & {$\fT$} & {AUC} \\
    \midrule
    1 & 5 & 0.890 & 2  & 0.705 & 31 & 0.654 \\
    2 & 3 & 0.889 & 15 & 0.702 & 30 & 0.654 \\
    3 & 6 & 0.889 & 5  & 0.701 & 5 & 0.654 \\
    4 & 8 & 0.888 & 23 & 0.701 & 3 & 0.653 \\
    \bottomrule
  \end{tabular}
\end{table}

\subsection{IIVV selection}
The best four performing IIVV sets using the SVM for all three classes tuned for the AUC are reported in Table~\ref{tab:Best4IIVV_reduction}. 
AUCs changed depending on the IIVV pattern set: \num{0.85} to \num{0.89} for \textit{Task 1}, \num{0.77} to \num{0.81} for \textit{Task 2}, and \num{0.76} to \num{0.79} for \textit{Task 3}.
These results demonstrate the influence of IIVV pattern selection on model classification performance.

\begin{table}[ht]
  \caption{Best four rankings for IIVV selection based on the AUC for the different tasks using the baseline SVM model.}
  \label{tab:Best4IIVV_reduction}
  \centering
  \resizebox{1.0\columnwidth}{!}{%
\begin{tabular}{@{}
    S[table-format=2]
    l
    S[table-format=1.3]
    l
    S[table-format=1.3]
    l
    S[table-format=1.3]
  @{}}
    \toprule
         & \multicolumn{2}{c}{\textit{Task 1}}
         & \multicolumn{2}{c}{\textit{Task 2}}
         & \multicolumn{2}{c}{\textit{Task 3}} \\
    \cmidrule(lr){2-3} \cmidrule(lr){4-5} \cmidrule(lr){6-7}
         {Rank} & {IIVV Mask} & {AUC} & {IIVV Mask} & {AUC} & {IIVV Mask} & {AUC} \\
    \midrule
 1 & \textit{Opp. med.} & 0.894 & \textit{Skip1 far}   & 0.806 & \textit{Skip1 far}   & 0.791 \\
 2 & \textit{Adj. far}  & 0.893 & \textit{Med. a+ ext.} & 0.802 & \textit{z-thresh} & 0.787\\
 3 & \textit{Skip1 far} & 0.889 & \textit{Skip1 med.} & 0.801 & \textit{Skip1 med.}  & 0.782 \\
 4 & \textit{Skip1 med.} & 0.889 & \textit{Opp. med.} & 0.792 & \textit{Opp med.} &  0.781 \\
    \bottomrule
  \end{tabular}
  }
\end{table}

\begin{table*}[ht]
  \caption{Baseline results for the default SVM, RF, and LR models with the full preprocessed impedance measurements. Metrics are average results from 5-fold LOPOCV with bold text denoting the top score.}
  \label{tab:BaselineSVM_RF_LR}
  \centering
\begin{tabular}{@{}
    S[table-format=2]
    S[table-format=1.2]
    S[table-format=1.2]
    S[table-format=1.2]
    S[table-format=1.2]
    S[table-format=1.2]
    S[table-format=1.2]
    S[table-format=1.2]
    S[table-format=1.2]
    S[table-format=1.2]
    S[table-format=1.2]
    S[table-format=1.2]
    S[table-format=1.2]
    S[table-format=1.2]
    S[table-format=1.2]
    S[table-format=1.2]
  @{}}
    \toprule
         & \multicolumn{5}{c}{\textit{Task 1}}
         & \multicolumn{5}{c}{\textit{Task 2}}
         & \multicolumn{5}{c}{\textit{Task 3}} \\
    \cmidrule(lr){2-6} 
    \cmidrule(lr){7-11} 
    \cmidrule(lr){12-16}
         {Model} 
         & {Acc} & {F1} & {AUC} & {Rec} & {Prec}
         & {Acc} & {F1} & {AUC} & {Rec} & {Prec}
         & {Acc} & {F1} & {AUC} & {Rec} & {Prec} \\
    \midrule
 SVM & \textbf{0.82} & \textbf{0.81} & 0.88 & 0.81 & \textbf{0.81}
     & 0.59 & 0.59 & 0.78 & 0.59 & 0.59
     & 0.59 & 0.59 & 0.79 & 0.59 & 0.59  \\
 LR  & 0.78 & 0.77 & 0.81 & 0.79 & 0.77
     & 0.58 & 0.58 & 0.75 & 0.58 & 0.58
     & 0.53 & 0.53 & 0.73 & 0.53 & 0.53 \\
 RF  & 0.81 & 0.81 & \textbf{0.89} & \textbf{0.83} & 0.80
     & \textbf{0.60} & \textbf{0.60} & \textbf{0.80} & \textbf{0.60} & \textbf{0.60}
     & \textbf{0.59} & \textbf{0.59} & \textbf{0.81} & \textbf{0.59} & \textbf{0.59} \\
    \bottomrule
\end{tabular}
\end{table*}

\begin{table*}[ht]
  \caption{Model parameter tuning results overview with the best achieved AUCs for the models SVM, RF, and LR and the tuning metric used. 
    The number of input parameters $N_{input}$ represents the number of input data points fed into the classification model. 
    The baseline AUC results are based on $N_{input}=\num{239568}$, using all IIVV and frequencies.
    For \textit{Tasks 2 and 3}, the micro-average metrics are presented.}
  \label{tab:results_summary}
  \centering
\begin{tabular}{@{}
    S[table-format=1] 
    c                 
    c                 
    S[table-format=2] 
    l                 
    S[table-format=5] 
    S[table-format=1.3] %
    S[table-format=1.3] %
    S[table-format=1.3] %
    S[table-format=1.3] %
    S[table-format=1.3] %
@{}}
\toprule
{\textit{Task}} & {Model} & {Tuning metric} & {Best $\fT$} & {IIVV mask} & {$N_{\text{params}}$}
& {Acc} & {F1} & {AUC} & {Recall} & {Precision} \\
\midrule
\textit{1} & LR & AUC & {8}  & \textit{Skip1 far}  & {1792}  & {0.80} & {0.79} & {0.90} & {0.77} & {0.81} \\
\textit{2} & LR  & Acc & {2} & \textit{Skip1 far} & {448}  & {0.62} & {0.62} & {0.82} & {0.62} & {0.62} \\
\textit{3} & RF  & AUC & {3} & \textit{Opp. medium}   & {792} & {0.60} & {0.60} & {0.83} & {0.60} & {0.60} \\
\bottomrule
\end{tabular}
\end{table*}

\subsection{Parameter Tuning}
Parameter optimization improved classifier performance across tasks as follows: 
\paragraph*{Task 1}
The best SVM model parameters were a \textit{poly} kernel and a regularization parameter \textit{C} set to \num{0.001}, which resulted in an AUC of \num{0.91}. 
The RF classifier was optimized to have \num{300} trees (estimators), a max depth of \num{8}, max features of \num{0.3}; this resulted in an AUC of \num{0.91}. 
For the LR classifier, the tuned model had a regularization parameter \textit{C} of \num{0.001} and \num{200} maximum iterations, resulting in an AUC of \num{0.91}. 

\paragraph*{Task 2}  
The best SVM model had a \textit{linear} kernel with regularization parameter \textit{C} at \num{0.001}, yielding an AUC of \num{0.80}.  
The RF classifier performed best with \num{200} trees, max depth of \num{8}, max features set to \num{0.7}, achieving an AUC of \num{0.78}.  
For LR, the best model had a regularization parameter \textit{C} of \num{0.001} and \num{2000} maximum iterations, resulting in an AUC of \num{0.79}.  

\paragraph*{Task 3}
For the multiclass classification, parameter tuning identified the SVM with a \textit{linear} kernel and regularization parameter \textit{C} at \num{0.01}.  
The RF classifier performed best with \num{800} trees, max depth \num{8}, max features \num{0.7}.  
LR was optimized with \textit{C} as \num{0.01} and \num{500} iterations.  
RF performed best in multiclass classification, achieving an AUC of \num{0.78}, compared with \num{0.77} for SVM and \num{0.77} for LR. 

\subsection{Classification}
Table \ref{tab:BaselineSVM_RF_LR} reports micro-averaged metrics for SVM, RF, and LR Baseline models for each task. 
Metrics on the classification performance of the best performing reduced dataset models are shown in Table \ref{tab:results_summary}. 
The optimized classifiers showed stable or improved discriminative performance compared to the Baseline results across all tasks.


For the binary \textit{Task 1}, LR tuned with AUC yielded the highest AUC of all models (\num{0.90}) using the \textit{Top 8} frequencies and the \textit{Skip1 far} IIVV patterns, requiring $N_{input}=\num{1792}$ (less than \SI{1}{\percent} of baseline).  
This model improved the Baseline LR model scores and achieved the highest AUC across all Baseline models, but showed just minor decrease in accuracy from the Baseline SVM and RF results. 
Models trained with all three scoring metrics in the three class multiclass problem of \textit{Task 2} improved the Baseline results. 
LR classification tuned with accuracy recorded the best AUC of \num{0.82} with the \textit{Top 2} frequencies and \textit{Skip1 far} IIVV pattern set. 
For \textit{Task 3}, RF tuned with AUC achieved the best performance using $\mathbf{f}_{\mathrm{T=5}}$ frequencies with \textit{Opp. medium} IIVV patterns ($N_{input}=\num{792}$), reaching an average AUC of \num{0.83}. 
Notably, the reduced model inputs for \textit{Task 3} achieved a higher micro-average AUC than the top model for \textit{Task 2} and significantly improved all classification metrics from Baseline models.
For full results and confusion matrices, see \cite{Thones_IIVV_Evaluation_2025}.

Overall, the results of this study demonstrate that a carefully selected subset of frequency components combined with targeted IIVV masks can sustain or enhance predictive performance.  
Models showed up to \SI{8}{\percent} improvements in AUC from the baseline results on average and typically used the least amount of input parameters. 

\begin{figure}
    \centering
    \begin{subfigure}[ht]{1\columnwidth}
        \captionsetup{position=above}
        \centering
        \includegraphics[width=0.9\columnwidth]{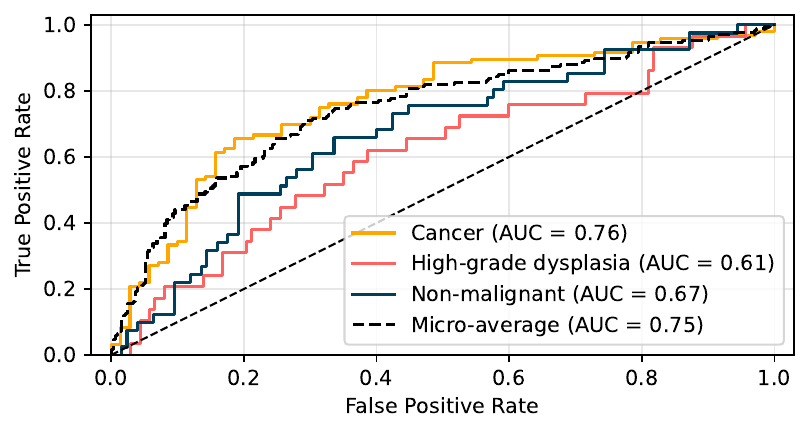}
        \caption{Baseline LR model}
        \label{fig:ROC_task2_baseline_lr}
    \end{subfigure}
    \begin{subfigure}[ht]{1\columnwidth}
        \captionsetup{position=above}
        \centering
        \includegraphics[width=0.9\columnwidth]{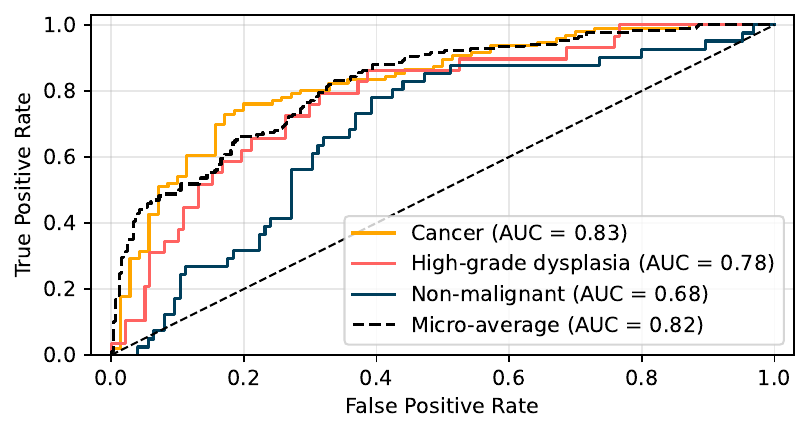}
        \caption{Tuned LR model}
        \label{fig:ROC_task2_lr_acc}
    \end{subfigure}
    \caption{Per-class and micro-average ROC performance for \textit{Task 2} using the LR model, tuned with accuracy as the scoring metric - results from fold one of 5 from the final cross validation.}
    \label{fig:ROC_task2_LR}
\end{figure}

Fig.~\ref{fig:ROC_task2_LR} shows ROC curves from the first of five folds for baseline and tuned LR models on \textit{Task 2}.
The tuned LR model demonstrated consistent improvements across all categories: cancer AUCs increased from \num{0.76} to \num{0.83}, high-grade dysplasia from \num{0.61} to \num{0.78}, non-malignant increased from \num{0.67} to \num{0.68}, and the micro-average AUC rose to \num{0.82}. 
This suggests that the parameter tuning helped the model capture previously underrepresented patterns.
The other ROC plots for baseline vs tuned models can be found in \cite{Thones_IIVV_Evaluation_2025}.

In addition to the demonstrated classification improvement and robustness, we evaluated prediction times for each model (run on a Supermicro A+ Server 4124GO-NART+ with dual AMD Epyc 7773X, 64 core, \SI{2.2}{\giga\hertz} and \SI{2}{\tera\byte} RAM).
The reduction of the input data dimension $N_{input}$ directly impacts the prediction time and thereby computational cost.
The largest cuts in computation costs occurred in the tuned SVM models.
SVMs are computationally expensive, particularly with higher input data dimensions.
Consequently, reducing the input dimensionality led to a substantial speedup.
LR scales linearly with the number of samples, which results in temporal and computational improvements when reducing the input size.
RF benefited less from dimensionality reduction, as their tree-splitting procedures remain relatively efficient.
Instead, using higher values for the hyperparameters (e.g., number of trees default: \num{100}, tuned: \num{800}) 
of the model, increased the computational costs compared to the baseline performance.
With fully trained models, a single prediction takes approximately \SI{0.01}{\second}. 

\section{Discussion}
\label{sec:discussion}

The dimension reduction framework developed in this study resulted in significant cuts in both the frequency and IIVV pattern domains for this oral cancer EIS dataset.
Using these reduced domains proved effective for classifying oral tissue samples in both binary and multiclass settings, with top parameters outperforming models trained on the complete dataset. 
This supports our hypothesis that selective inclusion of the most informative frequencies and patterns improves classification accuracy by removing noisy or less informative measurements that dilute impedance contrast. 
These findings directly address the two key barriers limiting clinical adoption of EIS for oral lesions: (1) reliably distinguishing OSCC and high-risk dysplastic lesions requiring intervention from low-risk lesions suitable for monitoring, and (2) providing clinicians with easy-to-interpret, actionable outputs to support confident biopsy decisions.
These results represent a significant advancement over our prior work \cite{lloyd_vivo_2025}, where classification relied exclusively on long across IIVV patterns and single-frequency threshold-based methods. 
In contrast, the current study incorporates a comprehensive range of IIVV patterns and employs optimized frequency selection combined with classical ML classifiers.
This approach not only improves classification accuracy but also allows for a more nuanced discrimination across multiple tissue classes, demonstrating the benefits of systematically expanding both pattern diversity and frequency utilization.
It also allows for easy interpretation of prediction outcomes for clinical use. 

\begin{figure*}
    \centering
    \footnotesize
\begin{tikzpicture}

\def\scale{1.0}
\def\h{0.65}
\def\dy{0.9}
\def\baseline{1} 
\def\tuned{0.3}
    \def\datameas{8}
    \def\resetprobe{1.5}
    \def\filtering{1.3}
    \def\total{\datameas+\filtering}
    \node[anchor=east] at (-0.3,\h/2) {Full};
    
    \draw[] (0,0) rectangle ++({\datameas*\scale},\h);
    \draw[]({\datameas*\scale},0) rectangle ++({\resetprobe*\scale},\h);
    \draw[]({(\datameas+\resetprobe)*\scale},0) rectangle ++({\filtering*\scale},\h);

    \draw[fill=Tuned]({(\datameas+\resetprobe+\filtering)*\scale},\h/2) rectangle ++({\tuned*\scale},\h/2);
    \draw[fill=Baseline]({(\datameas+\filtering+\resetprobe)*\scale},0) rectangle ++({\baseline*\scale},\h/2);
    \node[text width=4cm, align=center] at ({\datameas*\scale/2},\h/2) {Data collection ($\times3$ bursts) \par (\SI{\datameas}{\second})};
    \node[text width=2cm, align=center] at ({\datameas*\scale + \resetprobe*\scale/2},\h/2) {Reset probe \par (\SI{\resetprobe}{\second})};
    \node[text width=2cm, align=center] at ({\datameas*\scale + \resetprobe*\scale + \filtering*\scale/2},\h/2) {Filtering \par (\SI{\filtering}{\second})};

    \node[align=center] at ({(\datameas+\filtering+\resetprobe+\tuned/2)*\scale},3*\h/4) {+};
    \node[align=center] at ({(\datameas+\filtering+\resetprobe+\baseline/2)*\scale},\h/4) {+};
    
    \def\datameas{4}
    \def\filtering{1.15}
    \def\resetprobe{1.5}
    \node[anchor=east] at (-0.3,-\dy+\h/2) {Reduced};
    
    \draw[] (0,-\dy) rectangle ++({\datameas*\scale},\h);
    \draw[] ({\datameas*\scale},-\dy) rectangle ++({\resetprobe*\scale},\h);
    \draw[]({(\datameas+\resetprobe)*\scale},-\dy) rectangle ++({\filtering*\scale},\h);

    \draw[fill=Tuned]({(\datameas+\filtering+\resetprobe)*\scale},-\dy+\h/2) rectangle ++({\tuned*\scale},\h/2);
    \draw[fill=Baseline]({(\datameas+\filtering+\resetprobe)*\scale},-\dy) rectangle ++({\baseline*\scale},\h/2);
    
    \node[text width=4cm, align=center] at ({\datameas*\scale/2},-\dy+\h/2) {Data collection ($\times3$ bursts) \par (\SI{\datameas}{\second})};
    \node[text width=2cm, align=center] at ({\datameas*\scale + \resetprobe*\scale/2},-\dy+\h/2)
      {Reset probe \par (\SI{\resetprobe}{\second})};
    \node[text width=1.5cm, align=left, scale=0.95] at ({\datameas*\scale + \resetprobe*\scale + \resetprobe*\scale/2},-\dy + \h/2) {Filtering \par (\SI{\filtering}{\second})};

    \node[align=center] at ({(\datameas+\filtering+\resetprobe+\tuned/2)*\scale},-\dy +3*\h/4) {+};
    \node[align=center] at ({(\datameas+\filtering+\resetprobe+\baseline/2)*\scale},-\dy +\h/4) {+};
    
\draw[->] (0,-1.1) -- ({14.5},-1.1)
  node[below] {time in s};
\foreach \x in {0,1,...,13}
    {
      \draw ({\x*\scale},-1.1) -- ({\x*\scale},-1.25);
      \node[below] at ({\x*\scale},-1.25) {\x};
    }
\end{tikzpicture}
    \caption{Timings of the full clinical and computational workflow for one measurement. Data collection, with one measurement with 3 bursts recorded, filtering of the bursts, probe reset and the duration of the different classification models are shown with times in parentheses. The average baseline and tuned model times are shown in red and green, respectively.}
    \label{fig:timimgs_data_collection}
\end{figure*}
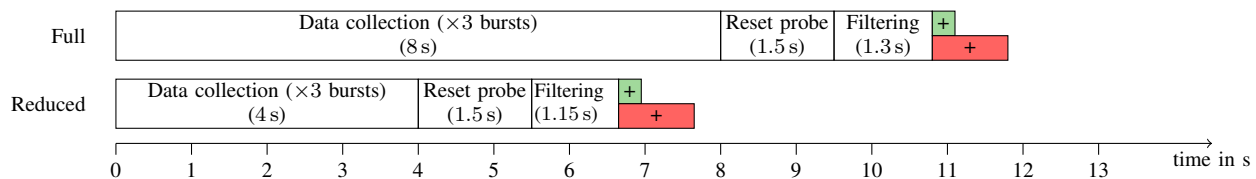

Classical ML models were selected due to their lower computational costs, robustness against overfitting, and ease of interpretation.
This design choice directly supports our second objective of providing clinicians with interpretable outputs; our feature reduction pipeline identifies exactly which frequencies and IIVV patterns offer the most critical information for cancer discrimination, enabling transparent insight into model decisions.
SVMs, in particular, are well-suited to high-dimensional spaces where the number of features exceeds the number of samples, as in this dataset. 
They are also memory-efficient and therefore well-positioned for clinical deployment in real-time analysis software. 
Traditional models such as RF, SVM, and LR often outperform deep neural networks (DNNs) or convolutional neural networks (CNNs) when data are limited or when interpretability and computational efficiency are priorities \cite{kourou2015machine, cortes1995support}.
Moreover, the computational resources needed to train and deploy CNNs or DNNs are significantly higher, a major limitation for real-time clinical applications on standard workstations or embedded hardware. 
We intentionally omitted CNNs or DNNs from our comparisons to avoid unfairly contrasting models with vastly different complexities, particularly as input dimensionality varies widely. 

For instance, Lin et al. \cite{lin_deep_2025} reported a CNN classifier distinguishing low-risk vs high-risk oral lesions with an AUC of 0.907.
Their low-risk group was predominantly healthy tissue (3 mild dysplasia readings from 2 patients), and their high-risk group primarily cancerous lesions (8 high-grade dysplasia readings from 4 patients).
Although their AUCs suggest strong performance, their study did not maintain strict patient-level separation between training and testing data, which has the potential to introduce data leakage and inflate results. 
Additionally, their smaller sample size may have exacerbated overfitting risks inherent to CNNs trained on limited data. 
Our healthy vs cancer binary classification (\textit{Task 1}) achieved comparable AUCs around 0.90 using classical ML models with rigorous patient splitting.
When classifying across all tissue types (\textit{Task 3}), AUCs for non-malignant (0.75) increased by 0.05, while high-grade dysplasia (0.65), cancer (0.88) and healthy (0.82) classes remained stable. This reflects the challenge of intermediate lesion discrimination.
\textit{Task 2} directly addresses our first objective by stratifying lesion samples into clinically actionable groups: those requiring immediate resection, those warranting clinical judgment regarding resection or monitoring, and those suitable for continued surveillance.
Retaining high cancer classification accuracy (0.88 AUC) in these more challenging settings demonstrates the value of our optimized multi-frequency, multi-pattern ML approach grounded in methodological rigor.

Our preliminary results in \cite{lloyd_vivo_2025}, which used only long across IIVV patterns and single-frequency threshold classifiers, reported AUCs of 0.85 for healthy vs cancer from impedance magnitude at \SI{160}{\hertz}, and 0.88 using permittivity at \SI{40}{\kilo\hertz}. 
\textit{Task 1} models in the current study outperform the best single-frequency permittivity classifier for \textit{Task 1} (healthy vs cancer), achieving AUCs of 0.90. 
However, the inclusion of dysplastic and non-malignant lesion groups in \textit{Task 2 and 3} represents a more complex and clinically relevant problem, resulting in lower overall AUCs compared to the simpler binary classification in \cite{lloyd_vivo_2025}. 
This emphasizes that the expanded use of multi-pattern, multi-frequency ML approaches enables classification across more challenging lesion types while retaining strong cancer discrimination.

Two IIVV masks consistently emerged as the most informative during parameter optimization: the \textit{Skip1 far} and the \textit{Opposite}-based masks.
The \textit{Skip1 far} set employs current drive electrodes with medium penetration depth (mean II distance: \SI{4.4}{\milli\meter}), whereas the \textit{Opposite medium} uses electrode pairs with deeper penetration (II distance: \SI{6.2}{\milli\meter}), see Table \ref{tab:masking_overview} for further detail.
Notably, the highest-ranking IIVV pattern sets from training did not always yield commensurately strong performance when combined with the reduced-frequency data sets. 
This discrepancy arises because sequential tuning implicitly assumes conditional independence between parameter groups, constraining the model to adapt along a single degree of freedom at a time.
The observed performance improvement and degradation upon combination reveals non-trivial interaction effects between the selected frequency subsets and IIVV masks. 
To mitigate this limitation, frequency and IIVV pattern sets were optimized independently to maintain tractable computational costs; jointly optimizing all combinations would increase the search space exponentially.
Future work could explore simultaneous tuning of both parameter groups to better capture these interactions.

Classification performance also varied considerably across IIVV pattern sets despite all measurements being derived from the same tissue samples.
This variation reflects the inherent heterogeneity of the underlying tissue: cancerous invasion depth is spatially non-uniform, and the degree and morphology of dysplasia can vary appreciably even within a region spanning only a few millimeters.
Because each IIVV pattern interrogates a distinct tissue volume and depth, each is differentially sensitive to this spatial heterogeneity.
Were the tissue perfectly homogeneous, all patterns would be expected to achieve comparable accuracy; the observed disparity therefore indicates that electrode geometry and tissue microstructural variability jointly limit diagnostic sensitivity.

Regarding frequency, \SI{100}{\hertz} and \SI{100}{\kilo\hertz} were the top two most important frequencies across all tasks, with \SI{1.3}{\kilo\hertz}, \SI{500}{\hertz}, and \SI{63}{\kilo\hertz} following closely. 
These results are similar to our preliminary results where the top performing frequencies for classification were usually in the low hertz to low  kilohertz range \cite{lloyd_vivo_2025}. 
Notably, \SI{100}{\kilo\hertz} was not a top frequency for any individual metric in \cite{lloyd_vivo_2025}. 
With the frequencies identified here, the extracellular electrical properties captured by low frequencies and the intracellular properties captured by the high frequencies can both be used for tissue discrimination. 
Frequencies from each of the four decades included in the frequency range placed in the top four frequencies, demonstrating the importance of capturing data across the 
dielectric tissue dispersions, which typically occur in the low kilohertz range for non-malignant oral mucosa \cite{scharfetter_low-frequency_2007, lloyd_vivo_2025}. 
These findings are consistent with the prior work from Veil et al. \cite{veil_electrical_2025} which employs feature reduction techniques to reduce the EIS frequencies from 50 to 5 for improved classification of bladder cancer from ex vivo tissue. 
Our work extends this concept specifically to oral cavity tissue, confirming that targeted reduction in both frequency and electrode pattern dimensions improves EIS diagnostic utility. 
Future work could focus on collecting impedance data at an even wider range of frequencies to ensure the full tissue dispersions for all types of oral lesions are captured. 

Current data acquisition time takes roughly \SI{8}{\second} to collect all IIVV patterns at all frequencies for three bursts.
Our findings suggest the number of frequencies can be drastically reduced to only five frequencies, resulting in a data acquisition time of \SI{4}{\second}.
Fig.~\ref{fig:timimgs_data_collection} demonstrates the timing on the full data collection and processing workflow for a single measurement (including three burst measurements). 
The filtering process was run on a laptop computer via MATLAB and is only marginally improved by the reduction in frequencies. 
The dataset reduction improves the full workflow by \SI{4}{\second}, which would have added up to nearly an hour of time saved if used on all three measurements of the 276 samples included in this study.
Our analysis also shows that while information from all eight current drive electrodes adds diagnostic value, acquisition need not include all \num{24} central voltage electrodes, which could alleviate hardware constraints for a custom EIT measurement system. 


Although this study represents the largest in vivo oral cavity EIS dataset to date, the number of dysplastic samples remains limited.
Increasing sample size, especially for underrepresented classes, would enhance the ability to train models on non-malignant and dysplastic lesions and improve classification performance, which is currently limited. 
Additionally, models in this study were trained using the impedance magnitude without the phase information. 
Introducing phase would have doubled the number of inputs to the models and required additional feature reduction steps for the phase values. 
Prospective clinical validation is necessary to confirm the efficacy of these models in new patient populations. 
Larger datasets would also enable the use of more complex models, such as deep learning architectures, once the number of samples approaches or exceeds the number of features.

Additionally, the current analysis was conducted offline. 
Given the promising results, implementing this algorithm for real-time oral lesion classification in prospective clinical studies is a feasible approach.
The 3D-printed nature of the EIS device ensures cost-effective manufacturing, facilitating further testing and deployment.
 If clinically implemented, this framework could provide dental clinicians with tissue-type classification probabilities to aid diagnostic decisions, directly fulfilling our objective of delivering actionable, easy-to-interpret outputs that support confident referral or biopsy recommendations.

In summary, frequency reduction results demonstrate that reliable discrimination between healthy and cancerous oral tissue is achievable with only a handful of top frequency components, paving the way for faster, simpler, and more efficient clinical measurements.
This work addresses both key barriers to clinical EIS adoption in the oral cavity: \textit{Task 1} confirms accurate cancer discrimination from healthy tissue, \textit{Task 2} enables clinically relevant stratification of lesions by intervention urgency, and the transparent, interpretable nature of our classical ML and feature reduction pipeline provides the foundation for actionable clinical decision support.

\section{Conclusion}
In this article, we proposed a novel dimension-reduction pipeline that significantly reduces input data while enhancing classification performance of EIS for diagnosing oral cancer and related lesions. 
Using optimized subsets of frequencies and IIVV patterns, parsimonious and interpretable ML models achieved high diagnostic accuracy, with AUCs up to \num{0.90} for healthy vs cancer classification and \num{0.83} for multiclass discrimination. 
Notably, classification performance for severe dysplasia and non-malignant lesions was lower in \textit{Tasks 2} and \textit{3}, highlighting the need for expanded datasets with more non-cancerous samples. 
Future work should focus on collecting larger, more diverse cohorts to improve classification of these challenging lesion types, ultimately supporting better monitoring and early detection. 
Overall, these results demonstrate the potential for streamlined, cost-effective EIS-based devices to provide rapid, non-invasive oral cancer diagnosis and surveillance in clinical settings.

\section{Data Availability}
Interested readers can contact the corresponding author to request access to the data. 

\bibliographystyle{ieeetr}
\bibliography{bibliography}

\end{document}